\documentclass[preprint,aps,prb,nofootinbib,floatfix,superscriptaddress]{revtex4-2}
\usepackage[utf8]{inputenc}
\usepackage{amsmath}    % need for subequations
\usepackage{graphicx}   % need for figures
\usepackage{verbatim}   % useful for program listings
\usepackage{color}      % use if color is used in text
\usepackage{epstopdf}   % use for side-by-side figures
\usepackage{upgreek}    %for non-italic greek letters (units)
\usepackage[hidelinks]{hyperref}   % use for hypertext links, including those to external documents and URLs
\usepackage{textcomp}   %Trademark symbol
\usepackage[uncertainty-mode = separate]{siunitx}
\usepackage{dcolumn}
\usepackage{comment}
\usepackage[capitalise]{cleveref}
\usepackage{mhchem}
\newcommand{\ie}{\texorpdfstring{\textit{i.e.}}{i.e.}}
\newcommand{\eg}{\texorpdfstring{\textit{e.g.}}{e.g.}}
\newcommand{\ea}{\texorpdfstring{\textit{et al.}}{et al.}}

\newcommand{\OOP}{\texorpdfstring{out-of-plane}{out-of-plane}}
\newcommand{\IP}{\texorpdfstring{in-plane}{in-plane}}
\newcommand{\CW}{\texorpdfstring{clockwise}{clockwise}}
\newcommand{\CCW}{\texorpdfstring{counter-clockwise}{counter-clockwise}}

\newcommand{\mumax}{\texorpdfstring{MuMax$^{3}$}{MuMax³}}

\DeclareMathOperator\erf{erf}

\newcommand{\eindhoven}{Department of Applied Physics, Eindhoven University of Technology, P.O. Box 513, 5600 MB Eindhoven, the Netherlands}

\newcommand{\nijmegen}{Radboud University Nijmegen, Institute for Molecules and Materials, 6525 AJ Nijmegen, the Netherlands}

\newcommand{\AFM}{NT-MDT BV, Sutton 11A, Apeldoorn, 7327AB, the Netherlands}

\begin{document}

\title{Facilitating electrical and laser-induced skyrmion nucleation with a dipolar-field enhanced effective DMI}

\author{Mark~C.\,H.~de~Jong}
\affiliation{\eindhoven}
\author{Dinar~Khusyainov}
\affiliation{\nijmegen}
\author{Julian~Hintermayr}
\affiliation{\eindhoven}
\author{Bart~Sanders}
\affiliation{\eindhoven}
\author{Dmitry Kozodaev}
\affiliation{\AFM}
\author{Aleksei~V.~Kimel}
\affiliation{\nijmegen}
\author{Bert~Koopmans}
\affiliation{\eindhoven}
\author{Theo~H.\,M.~Rasing}
\affiliation{\nijmegen}
\author{Reinoud~Lavrijsen}
\email{r.lavrijsen@tue.nl}
\affiliation{\eindhoven}

\date{\today}

\begin{abstract}
We demonstrate experimentally how the nucleation of skyrmions in an \ce{Ir}, \ce{Co}, and \ce{Pt} based magnetic multilayer is affected by introducing a layer dependent sign for the Dzyaloshinskii-Moriya interaction (DMI). In one stack, the bottom half of the stack is given a positive DMI and the top half a negative DMI, and as a result, the in-plane component of the dipolar field is aligned parallel to the effective field of the DMI in every layer, enhancing the effective DMI. We show that this enhanced DMI facilitates the nucleation and stability of skyrmions using both current-driven and laser-induced skyrmion nucleation. In the devices with an enhanced effective DMI, the density of nucleated skyrmions is greater by up to a factor $\sim$20 and skyrmions can be observed in stronger magnetic fields---suggesting that their stability is also improved. These results  show that skyrmion nucleation depends strongly on the magnitude of the effective DMI in a magnetic multilayer and that the dipolar field within such a multilayer presents an effective route towards controlling the effective DMI, and thereby, the nucleation of chiral magnetic textures. 
\end{abstract}

\maketitle

\section{Introduction}
A key interaction required for the stability of chiral magnetic textures such as skyrmions is the Dzyaloshinskii-Moriya interaction (DMI)~\cite{Wiesendanger2016,Fert2017,Everschor-Sitte2018,Zhang2020,Tokura2021}.
In magnetic multilayers, the interfacial DMI stabilizes chiral Néel-type domain walls and skyrmions over their Bloch-type counterparts. This chirality is vital for devices that make use of skyrmions as information carriers, since it ensures that all skyrmions in the device experience the same forces when a current is sent through the system~\cite{Tomasello2014, Zhang2020}---which enables shifting data through the device. Additionally, the DMI lowers the domain-wall energy density, which increases the stability of the skyrmion texture~\cite{Soumyanarayanan2017,Buettner2018}. Therefore, control over the DMI strength---and in particular, increasing the DMI strength---is of great technological and academic interest.

In a magnetic multilayer, however, the DMI is not the only interaction that can influence the chirality of domain walls and skyrmions~\cite{Lucassen2019,Meijer2020}. In systems where the thickness is larger than the exchange length and with a large magnetic moment, the dipolar interaction between the \OOP{} magnetization in the domains and the \IP{} magnetization inside the domain walls can be significant (compared to the DMI)~\cite{HubertOrigin1998,Lemesh2018a}, resulting in the formation of Néel caps~\cite{Legrand2018,Dovzhenko2018,Lucassen2019}. In the top of the stack, a \CW{} chirality is energetically favored and conversely, a \CCW{} chirality is favored in the bottom half of the stack. Depending on the stack geometry, magnetic parameters, and the scheme that is used to inject a spin-current into the multilayer, the dipolar interaction can significantly affect the current-driven motion of chiral magnetic textures~\cite{LegrandPhD2019} and is therefore often seen as a negative effect. Avoiding this effect is however nontrivial, as the dipolar field generated by the domains-—while detrimental to current-driven motion—-is essential for the high stability and small size of skyrmions~\cite{Moreau2016,Woo2016,Boulle2016,Soumyanarayanan2017}.

Here, we utilize this internal dipolar field in combination with opposite stacking orders of the layers in the top and bottom half of the stack, to either enhance or reduce the effective DMI of the system~\cite{Hrabec2017,Lucassen2020a}. By choosing the appropriate sign of the interfacial DMI for repeats in the bottom and top half,  a stack can be created where the DMI and the dipolar field stabilize the same (opposite) chirality in all layers, effectively enhancing (reducing) the DMI. Using this method we can study the effect of significant differences in the effective DMI of the stack on the nucleation of skyrmions in isolation, which has hitherto not been done in skyrmion nucleation experiments. This is because conventional methods of varying the magnetic parameters, such as changing the thickness of the magnetic layers (see e.g. \cite{Soumyanarayanan2017}), ion irradiation with either Ga$^{+}$ or He$^{+}$ ions (see e.g. \cite{Lemesh2018a,deJong2023,Kern2022}), or changing the stack composition \cite{Soumyanarayanan2017} almost always lead to changes in multiple magnetic parameters. In Ref. \cite{Lucassen2020a}, we showed that this dipolar induced change in the effective DMI can be measured and we examined its effect on the skyrmion energy using micromagnetic simulations. Here, we experimentally examine the effect on skyrmion nucleation and show that skyrmion nucleation is facilitated in the samples with an enhanced effective DMI when compared to a reduced effective DMI. Using the stacking orders described above, we fabricated two samples with identical (average) saturation magnetization and uniaxial anisotropy strength, but an effective DMI that is different by almost a factor three. We then use both nanosecond-current-pulse-driven and femtosecond-laser-pulse-induced skyrmion nucleation to study the effect of changes in the effective DMI strength on the skyrmion nucleation. Irrespective of the nucleation method, we find that the change in the effective DMI has a significant effect on the density and stability of nucleated skyrmions, while the threshold current and threshold fluence seem only weakly affected. Interestingly, the effect of the dipolar enhanced and reduced effective DMI on both the current-driven and laser-induced nucleation is qualitatively and quantitatively similar, even though the timescales on which the systems are excited are very different. These results not only show that the internal dipolar field can be exploited to facilitate the nucleation of chiral magnetic textures, but they also present a first insight into the dependence of laser-induced skyrmion nucleation on the internal dipolar interactions in a magnetic multilayer and a possible route towards measuring this dependence.

\section{Samples and Methods}
The samples have the following double trilayer structure as building blocks: [\ce{Pt}$(1)\,|\,$\ce{Co}$(1)\,|\,$\ce{Ir}$(1)$] and [\ce{Ir}$(1)\,|\,$\ce{Co}$(1)\,|\,$\ce{Pt}$(1)$], with the numbers in brackets indicating the thickness in \unit{nm}. The former (latter) trilayer has a positive (negative) intrinsic DMI and therefore stabilizes \CCW{} (\CW{}) Néel-type domain walls~\cite{Legrand2018}. 

\begin{figure}
\includegraphics{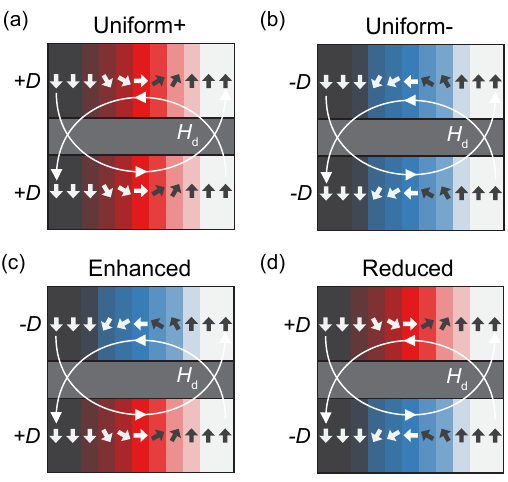}
\caption{
Sketches of the different multilayer structures.
(a) Uniform$+$ stack. All the layers in this stack have a positive DMI sign. (b) Uniform$-$ stack. Same as (a), but with the opposite DMI sign. (c) Enhanced stack. Here the magnetic layers in the top half of the stack have a negative DMI and the layers in the bottom half a positive DMI, as a result the DMI effective field and the dipolar field are aligned in all layers. (d) Reduced stack. Opposite DMI signs compared to the Enhanced stack in (c), now the DMI effective field and the dipolar field are antiparallel in all layers. Thick arrows represent magnetization and follow the DMI effective field, the long white arrows represent the dipolar field $\mu_{0} H_{\textrm{d}}$. The top and bottom half of each schematic represent multiple repeats with the same layer order.
}
\label{fig:Figure_1}
\end{figure}

In \cref{fig:Figure_1}~(a) and (b), we show a sketch of the structure of a typical magnetic multilayer. The bottom and top half of the stack contain the same trilayer-building block and therefore, have the same intrinsic DMI sign [(a) positive D, (b) negative D]. Consequently, the DMI in the top and bottom half of each stack stabilizes a uniform domain-wall chirality in the multilayer (thick arrows). However, in both stacks the dipolar field (white arrow labeled $H_{\textrm{d}}$) originating from the two domains (light and dark gray) is anti-parallel to the magnetization in the domain walls in either the top, or the bottom half  of the stack. This reduces the effective DMI in this part of the multilayer and---when the number of layers is large or the intrinsic DMI is weak---can give rise to heterochiral domain walls~\cite{Legrand2018,Dovzhenko2018,Lemesh2018a}. In \cref{fig:Figure_1}~(c), we show that introducing an opposite intrinsic DMI sign for the two halves of the multilayer can result in a DMI effective field that is parallel to the dipolar field everywhere in the multilayer, thus enhancing the effective DMI of the entire stack with respect to the intrinsic DMI strength of the building blocks~\cite{Lucassen2020a}.Additionaly, as we show in \cref{fig:Figure_1}~(c), the same mechanism can be used to fabricate a multilayer where the dipolar field and DMI effective field are everywhere antiparallel, reducing the effective DMI. Thus, by changing the stacking order of the Pt, Co, and Ir-based building blocks, the effective DMI strength of a multilayer can be controlled~\cite{Lucassen2020a}. These four configurations will be refered to as: (a) Uniform+, (b) Uniform-, (c) Enhanced stack, and (d) Reduced stack in the remainder of this article.

\clearpage
To this end we fabricate four magnetic multilayer structures consisting of the following stacks:
\begin{table}[htbp]
    \small
    \begin{tabular}{ @{}l l }
        \textbf{Uniform+:} & ${\overbrace{\mathrm{[\ce{Pt}}(1)\,|\,\mathrm{\ce{Co}}(1)\,|\,\mathrm{\ce{Ir}}(1)\mathrm{]}}^{+D}}_{\times 6}$, \\
        \\ 
        \textbf{Uniform-:} & ${\overbrace{\mathrm{[\ce{Ir}}(1)\,|\,\mathrm{\ce{Co}}(1)\,|\,\mathrm{\ce{Pt}}(1)\mathrm{]}}^{-D}}_{\times 6}$, \\
        \\
        \textbf{Enhanced:} & ${\overbrace{\mathrm{[\ce{Pt}}(1)\,|\,\mathrm{\ce{Co}}(1)\,|\,\mathrm{\ce{Ir}}(1)\mathrm{]}}^{+D}}_{\times \textrm{3}}$~$|$~${\overbrace{\mathrm{[\ce{Ir}}(1)\,|\,\mathrm{\ce{Co}}(1)\,|\,\mathrm{\ce{Pt}}(1)\mathrm{]}}^{-D}}_{\times \textrm{3}}$, \\
        \\
        \textbf{Reduced:} & ${\overbrace{\mathrm{[\ce{Ir}}(1)\,|\,\mathrm{\ce{Co}}(1)\,|\,\mathrm{\ce{Pt}}(1)\mathrm{]}}^{-D}}_{\times \textrm{3}}$~$|$~${\overbrace{\mathrm{[\ce{Pt}}(1)\,|\,\mathrm{\ce{Co}}(1)\,|\,\mathrm{\ce{Ir}}(1)\mathrm{]}}^{+D}}_{\times \textrm{3}}$.
    \end{tabular}
\end{table}

The substrate and seed layers for all stacks are \ce{Si}$\,|\,$\ce{SiO2}$(100)\,||\,$\ce{Ta}$(4)\,|\,$\ce{Pt}$(2)$ and the capping layer is either \ce{Ta}$(4)$ (characterization and laser-induced nucleation) or \ce{Pt}$(3)$ (current-driven nucleation). All samples are grown using DC magnetron sputtering in an \ce{Ar} atmosphere with a pressure of $p = $ \qty{2e-3}{\milli\bar} using a vacuum system with a base pressure better than $p = $ \qty{5e-9}{\milli\bar}. Using electron beam lithography and a standard lift-off procedure we pattern the samples into nucleation devices with large contact pads for wirebonding [$150 \times 100$ \unit{\micro\metre\squared}] and a narrow current line [$7 \times 1.6$ \unit{\micro\metre\squared}] (described in detail in Ref. \cite{deJong2023}).

Magnetometry measurements of the saturation magnetization and the effective anisotropy, as well as demagnetization of the samples were performed at room temperature in a Quantum Design MPMS3. To nucleate skyrmions we employ two custom-built magnetic force microscopy (MFM) setups. The current-driven nucleation experiments and the characterization of the DMI are performed using a Br{\"u}ker Dimension Edge atomic force microscope (AFM). A custom sample holder enables manipulation of the magnetic state of the multilayers using single \qty{50}{ns} current pulses from an Agilent 33250A \qty{80}{MHz} arbitrary waveform generator, as well as the simultaneous application of a magnetic field up to $\mu_{0} H_{\textrm{z}} = $ \qty{500}{mT} during measurements. The pulse voltage is measured using an Agilent Infiniium DSO80604B oscilloscope connected in series and the current through the device is calculated by dividing the measured pulse voltage by the resistance of the oscilloscope (\qty{50}{\ohm})~\cite{Lemesh2018}. Finally, the current density is calculated using the assumption that the current is uniformly distributed throughout the entire multilayer stack.

The laser-induced nucleation experiments are performed using an NT-MDT NTEGRA AFM. The scanhead of this AFM can be removed to allow access to the sample for a single \qty{70}{fs} laser pulse with a central wavelength of \qty{800}{nm}, picked from the \qty{1}{kHz} pulse train generated by a Spectra-Physics Spitfire Ace laser amplifier system using a Pockel's cell. After the application of such a pulse, the scanhead can be placed back over the sample, which results in only a small displacement of the tip position on the order of \qty{1}{\micro\metre}. Hence, the same position can be measured before and after the application of a laser pulse with relative ease. The sharp edges of the devices are used to position the laser spot and to measure its full width at half maximum (FWHM). The laser-driven nucleation is then performed in the contact pads, away from the edges of the device. Since the laser spot is much larger than the area of the MFM scan, we assume that the fluence of the laser is equal to the peak fluence of the laser spot, everywhere in the MFM scan. The peak fluence is calculated by measuring the laser power, which is first converted into energy per pulse using the repetition rate of the laser and then into fluence using the measured FWHM of the spot, assuming a 2D-Gaussian beam profile. An \OOP{} magnetic field can be applied to the sample during skyrmion nucleation, using an electromagnet installed in the sample holder. When MFM images are taken with this setup, the magnetic field is switched off to improve the signal-to-noise ratio.

For these experiments we use two types of custom-coated MFM cantilevers fabricated from NANOSENSORS PPP-FMR AFM cantilevers. A magnetic coating is applied to these cantilevers using the same sputter deposition system used to grow the samples and consists of either $||\,$\ce{Ta}$(4)\,|\,$\ce{Co}$(5)\,|\,$\ce{Ta}$(5)$ or $||\,$\ce{Ta}$(4)\,|\,$\ce{Co}$(7.5)\,|\,$\ce{Ta}$(5)$. The two different coatings are used since we always try to maximize the magnetic moment of the tip, but observed that domain walls in the reduced stack are influenced by the tip magnetization if a coating of \ce{Co}$(7.5)$ is used. In the case of the laser-induced nucleation experiments, the \ce{Co}$(5)$ coating was used for all stacks.

\section{Results}
\subsection{Sample characterization}
\textit{Saturation magnetization and uniaxial anisotropy} -  We measured both the \IP{} and \OOP{} hysteresis loops using SQUID-VSM, which are shown in Section SI of the Supplemental Material for all four stacks. In \cref{fig:Figure_2}~(a), we plot the measured saturation magnetization $M_{\textrm{s}}$. The uniform$+$ and uniform$-$ stacks are found to have the lowest and highest saturation magnetization, respectively, which is consistent with other reports on \ce{Pt}$\,|\,$\ce{Co}$\,|\,$\ce{Ir} and \ce{Ir}$\,|\,$\ce{Co}$\,|\,$\ce{Pt} multilayers~\cite{Legrand2018}.

\begin{figure}[t]
\includegraphics{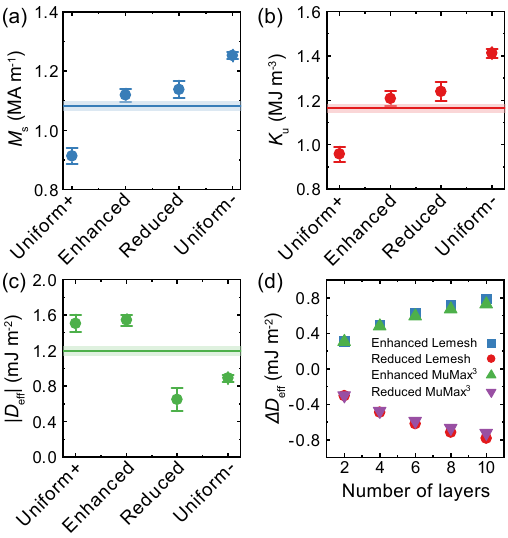}
\caption{
Characterization of the different magnetic multilayers. (a) The saturation magnetization $M_{\textrm{s}}$ and (b) uniaxial anisotropy $K_{\textrm{u}}$ measured using SQUID-VSM. (c) The effective DMI strength $\lvert D_{\textrm{eff}} \rvert$ determined by first measuring the equilibrium domain width and then using a model for the expected equilibrium domain width by Lemesh \textit{et al.}~\cite{Lemesh2017} to calculate $\lvert D_{\textrm{eff}} \rvert$. The horizontal lines in (a) - (c) show the average of the values measured for the two uniform stacks. (d) The expected contribution of the dipolar field to the effective DMI calculated using \mumax{} simulations (green and purple) and an analytical model by Lemesh \textit{et al.}~\cite{Lemesh2018a} (blue and red).
}
\label{fig:Figure_2}
\end{figure}

Additionally, we extract the effective anisotropy $K_{\textrm{eff}}$ from the SQUID-VSM hysteresis loops using the area method~\cite{Johnson1996}. We convert this value into the uniaxial anisotropy $K_{\textrm{u}}$ using
\begin{equation}
    K_{\textrm{eff}} = K_{\textrm{u}} - \frac{1}{2} \mu_{0} M_{\textrm{s}}^{2},
\end{equation}
\noindent and plot this value in \cref{fig:Figure_2}~(b). Since the measured effective anisotropy of the uniform$+$ and uniform$-$ stacks are similar, the trend in $K_{\textrm{u}}$ follows that of $M_{\textrm{s}}^{2}$ for these two stacks.

We will now consider the enhanced and reduced stack in the context of the two uniform stacks. Half of the layers in the enhanced and the reduced stack correspond to the stacking order of the uniform$+$ stack and the other half of those in the uniform$-$ stack. In Ref. \cite{Lucassen2020a}, we showed that the value of the saturation magnetization and the uniaxial anisotropy obtained from SQUID-VSM measurements for the enhanced and reduced stack can be understood as the average value of the two uniform stacks. Essentially, we assume that each layer in the multilayer contributes equally to the signal measured by the SQUID-VSM, and that each half of the enhanced and reduced stack has an $M_{\textrm{s}}$ and $K_{\textrm{u}}$ that is close to those found for the corresponding uniform stacks. Indeed, the measured values for $M_{\textrm{s}}$ [\cref{fig:Figure_2}~(a)] and $K_{\textrm{u}}$ [\cref{fig:Figure_2}~(b)] for the enhanced and reduced stacks are close to the average value for the two uniform stacks represented by the colored horizontal lines.The agreement between the values observed for the enhanced and reduced stacks with the average value of the two uniform stacks is not as good as that observed in Ref. \cite{Lucassen2020a}, We expect that this is because the number of layers in the multilayer was increased from four to six and that this slightly increased the cumulative effect of the bottom layers on the growth of the top layers.

\textit{The effective DMI} - We determined the effective DMI $\lvert D_{\textrm{eff}} \rvert$ using a model by Lemesh \ea{} \cite{Lemesh2017}. This model is used to predict the equilibrium domains size for a multilayer stack, if given the exchange stiffness $A$, the saturation magnetization $M_{\textrm{s}}$, the uniaxial anisotropy $K_{\textrm{u}}$, and the DMI strength $D$. By measuring the equilibrium domain size instead, it can also be used to determine $D$ if the other parameters are known~\cite{Lemesh2017, Agrawal2019} [see Ref. \cite{DeJong2022} for a detailed description of our implementation of this method].The results of this analysis are plotted in \cref{fig:Figure_2}~(c) for all four stacks. In Section SII of the Supplemental Material, we show MFM scans of the demagnetized domain states that are used to determine the equilibrium domain width. Contrary to the results for the saturation magnetization and the uniaxial anisotropy, the measured effective DMI for the enhanced and the reduced stack does not correspond to the average value of the two uniform stacks (horizontal green line in \cref{fig:Figure_2}~(c)). Instead, the DMI in the enhanced stack is found to be significantly larger than this average value (by \qty{0.34(8)}{mJ.m^{-2}}) and the DMI in the reduced stack is found to be significantly lower (by \qty{0.6(1)}{mJ.m^{-2}}), in accordance with our expectation based on \cref{fig:Figure_1}~(c) and (d).\footnote{A hypothesis for the difference in magnitude is given in Section SV of the Supplemental Material} The effective DMI in the enhanced stack is found to be a factor 2.5 greater than in the reduced stack. Despite this large difference in the effective DMI, the measured saturation magnetization and uniaxial anisotropy in both stacks are the same within our experimental uncertainty. This is important because this enables us to study the effect of the enhanced and reduced effective DMI on the nucleation of skyrmions directly, without having to account for changes in the nucleation due to the other material parameters.

\textit{Modelling} - We conclude this section with the results of \mumax{}~\cite{Vansteenkiste2014} simulations and an analytical model~\cite{Lemesh2018a}, the details of both models can be found in Section SV of the Supplemental Material. In the \mumax{} model we calculate the domain-wall energy of two multilayer stacks consisting of $N$ layers that are only coupled by the dipole interaction. The magnetic parameters for each layer in the simulation correspond to the average material parameters represented by the horizontal lines in \cref{fig:Figure_2}, except the DMI which we vary between $\lvert D \rvert = $ \qty{1}{mJ.m^{-2}} and $\lvert D \rvert = $ \qty{3}{mJ.m^{-2}}. The only difference between the two systems is the sign of the DMI in the top and bottom half, which is either: Enhanced $(D_{\textrm{Top}} < 0 \: ; \: D_{\textrm{Bottom}} > 0)$, or Reduced $(D_{\textrm{Top}} > 0 \: ; \: D_{\textrm{Bottom}} < 0)$. We then determine for which DMI strengths the two systems have the same domain-wall energy, the difference in DMI that is needed to obtain the same domain-wall energy is two times the gain/loss in effective DMI due to the dipolar field from the domains. In \cref{fig:Figure_2}~(d), we plot the obtained increase and decrease in the effective DMI for the enhanced and reduced stacks in green and purple, respectively. In addition, we also show the results from a model of the effective DMI due to the dipolar field by Lemesh \ea{}~\cite{Lemesh2018a} in \cref{fig:Figure_2}~(d) in blue and red. These results are in excellent agreement with the simulations performed using \mumax{}. The stack with $N = 6$ corresponds to the experiment and here we find a predicted change in the effective DMI of $\lvert \Delta D \rvert = $ \qty{0.6}{mJ.m^{-2}}. The simulated difference between the effective DMI of the enhanced and the reduced stack matches the measured difference ($\lvert \Delta D_{\textrm{Meas}} \rvert = $ \qty{0.47(6)}{mJ.m^{-2}}) relatively well, which confirms the dipolar origin of the difference in the effective DMI between the enhanced and reduced stack.

\subsection{Current-driven skyrmion nucleation}
The experimental procedure followed during the nucleation experiments is as follows: First, we saturate the magnetization in the device in the negative $z$-direction (\OOP{}, $\mu_{0}H_{\textrm{z}} < $ \qty{-150}{mT}). Then, a small bias field is applied is the positive $z$-direction and a single current (laser) pulse is sent to the device. Subsequently, an MFM scan is taken to record the magnetization state in the device.
In the case of the current-driven nucleation, the bias field is applied while the MFM scan is taken, while in the case of the laser-induced nucleation it had to be turned of before scanning to get a good signal-to-noise ratio.

\begin{figure*}[t]
\includegraphics[width=\textwidth]{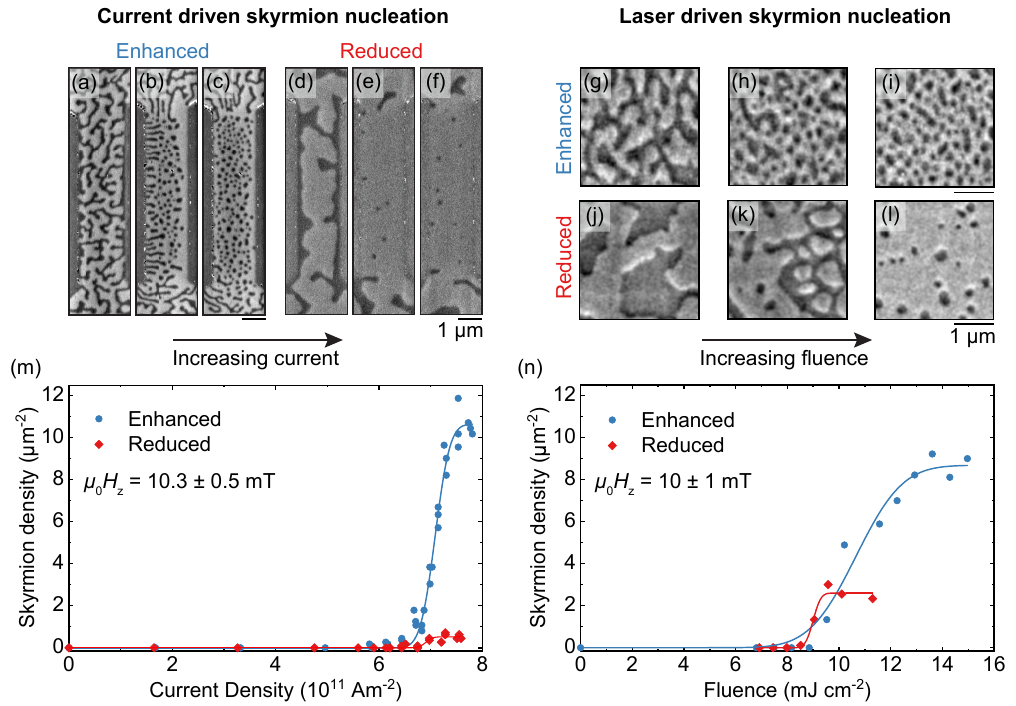}
\caption{
Effect of the enhanced and reduced effective DMI on the current- and laser-induced nucleation of skyrmions. In (a) - (l) we show MFM scans of the magnetization in the devices fabricated from the enhanced and reduced multilayer stack after applying a single nucleation pulse. (a) - (c): Enhanced stack after current-induced nucleation, the current densities are: $J = $ \qty{0}{A.m^{-2}}, $J = $ \qty{7.14e11}{A.m^{-2}}, and $J = $ \qty{7.77e11}{A.m^{-2}}, respectively. (d) - (f): Reduced stack after current-driven nucleation, the current densities are: $J = $ \qty{0}{A.m^{-2}}, $J = $ \qty{6.97e11}{A.m^{-2}}, and $J = $ \qty{7.55}{A.m^{-2}}. In (a) through (f) the current flows from top to bottom. (g) - (i): Enhanced stack after laser-induced nucleation, the fluences are: $F = $ \qty{8.9}{mJ.cm^{-2}}, $F = $ \qty{11.6}{mJ.cm^{-2}}, and $F = $ \qty{15.0}{mJ.cm^{-2}}, respectively. (j) - (l), Reduced stack after laser-induced nucleation, the fluences are: $F = $ \qty{6.9}{mJ.cm^{-2}}, $F = $ \qty{9.1}{mJ.cm^{-2}}, and $F = $ \qty{10.1}{mJ.cm^{-2}}. In (m) and (n), we summarize the results from the current-driven and laser-induced skyrmion nucleation, respectively. For each current density we plot the average skyrmion density in the scanned area, for both the enhanced (blue dots) and reduced stack (red diamonds). The solids lines in (m) and (n) are fits to the data using \cref{eq:error_function}.
}
\label{fig:Figure_3}
\end{figure*}

We will discuss the current-driven skyrmion nucleation experiments first. In \cref{fig:Figure_3}~(a) - (c), we show three MFM scans obtained after sending three different current (this includes $J = $ \qty{0}{A.m^{-2}})  densities through the device. For all three scans, the bias field is set to $\mu_{0} H_{\textrm{Bias}} = $ \qty{10.3(5)}{\milli\tesla}, along the positive $z$-direction, and the current densities during the pulses are $J = $ \qty{0}{A.m^{-2}}, $J = $ \qty{7.14e11}{A.m^{-2}}, and $J = $ \qty{7.77e11}{A.m^{-2}}, respectively. When no current is applied [\cref{fig:Figure_3}~(a)], the magnetization in the device is in a demagnetized state, the domain that is parallel to the applied field (light gray) is slightly bigger than the antiparallel domain. In \cref{fig:Figure_3}~(c), we show the magnetization after the highest current density that can be reached by our pulse generator is sent through the device. Here, the magnetization inside the narrow strip is fully in the skyrmion state with $117$ skyrmions in the strip. The average skyrmion diameter is found to be $d_{\textrm{sk}} = $\qty{135(3)}{nm}, see Section SIII of the Supplemental Materials for the details. The scan that is shown in \cref{fig:Figure_3}~(b) depicts the magnetization state when an intermediate current density is applied. Here, skyrmions only appear on the right side of the strip and worm-like domains fill the left half of the strip.\footnote{The orientation of these worm-like domains is perpendicular to the current flow. This has been observed before and a phenomenological explanation can be found in Ref. \cite{Lemesh2018}.} Since the direction of the current flow is top-to-bottom (along negative $y$), the Oersted field generated by the current pulse is parallel to the bias field in the right half of the stack. This increase in field decreases the stability of the antiparallel dark gray domains. If the polarity of the current is reversed the skyrmions nucleate on the other side of the strip, because the Oersted field changes sign.

We also repeated the nucleation experiment for a device fabricated from the reduced stack, with the same bias field applied. In \cref{fig:Figure_3}~(g) - (i), we show three MFM scans that were taken on a device fabricated from the reduced stack for zero current, an intermediate current ($J = $ \qty{6.97e11}{A.m^{-2}}) and a current above the nucleation threshold ($J = $ \qty{7.55e11}{A.m^{-2}}), respectively. The density of skyrmions in the device after the nucleation pulse is applied is much lower compared to the devices fabricated from the enhanced stack, with only a few isolated skyrmions present in the strip after nucleation. For the scans above the threshold current we can determine the average skyrmion diameter and find $d_{\textrm{sk}} = $ \qty{184(7)}{nm}, slightly larger than in the case of the enhanced stack.
A lower effective DMI is expected to decrease the skyrmion size~\cite{Buettner2018}, because the domain-wall energy density increases, which is not observed. We hypothesize that, in the case of the enhanced stack, the high skyrmion density in combination with the narrow strip prevents the skyrmions from expanding to the radius expected for isolated skyrmions.

We used the same skyrmion counting procedure that was used in Ref. \cite{deJong2023} to determine the number of skyrmions in the device for both the enhanced and reduced stacks. To facilitate the comparison with the laser-induced skyrmion nucleation later, we convert the number of skyrmions into the skyrmion density by dividing by the strip area: $7 \times 1.6$ \unit{\micro\metre\squared}. In \cref{fig:Figure_3}~(m), we plot the skyrmion density $\rho_{\textrm{Sk}}$ as a function of the applied current density, the solid lines are fits to the data using the following equation~\cite{deJong2023},
\begin{equation}\label{eq:error_function}
    \rho_{\textrm{Sk}} = \frac{1}{2} \rho_{\textrm{Sk,sat}} \Big[ 1 + \erf \Big( \frac{J - J_{\textrm{c}}}{\sqrt{2} \, \sigma} \Big) \Big],
\end{equation}
\noindent where $J_{\textrm{c}}$ is the threshold current density and $\sigma$ the width of the transition region. The maximum skyrmion density that is reached is $\rho_{\textrm{Sk,sat}}^{\textrm{Enh}} = $ \qty{10.6(1)}{\per\micro\metre\squared} and $\rho_{\textrm{Sk,sat}}^{\textrm{Red}} = $ \qty{0.52(1)}{\per\micro\metre\squared} for the enhanced and reduced stacks, respectively. Hence, the number of skyrmions after nucleation in the enhanced stack is greater by a factor of 20 compared to the reduced stack. Because the measured saturation magnetization and uniaxial anisotropy of both stacks are the same within our experimental uncertainty, and the difference in resistance is only $3\%$~\footnote{Using a $2$-wire geometry, the resistance of the devices fabricated from the enhanced and reduced stacks are measured to be $R_{\textrm{Enh}} = $ \qty{168.2}{\ohm} and $R_{\textrm{Red}} = $ \qty{173.3}{\ohm}, respectively.}, we attribute this difference in skyrmion density to the dipolar enhancement and reduction of the effective DMI of our multilayer stacks.

Surprisingly, considering the results presented in Ref. \cite{deJong2023}, there is almost no difference in the threshold current that we observe for the two stacks. We find $J_{\textrm{c}} = $ \qty{7.09(1)e11}{A.m^{-2}} and $J_{\textrm{c}} = $ \qty{6.90(2)e11}{A.m^{-2}} for the enhanced and reduced stacks, respectively. Hence, the skyrmion density appears to be much more sensitive to the increase and decrease in effective DMI than the threshold current. Before we discuss these results further, we will present the results from the laser-induced skyrmion nucleation in the same two material stacks. 

\subsection{Laser-induced skyrmion nucleation}
MFM images obtained after laser-induced nucleation as a function of fluence for both samples are shown in \cref{fig:Figure_3}~(g) through (l). Additionally, we plot the skyrmion density as a function of fluence in \cref{fig:Figure_3}~(n). The nucleation of a skyrmion with a laser pulse shows a threshold behavior as a function of fluence, qualitatively similar to the current-driven nucleation experiment. Far below the threshold, the sample continues to have the labyrinth domain state after the laser pulse has hit the sample. The threshold fluences for both samples are found to be $F_{\textrm{c}} = $ \qty{9.0(2)}{mJ.cm^{-2}} and $F_{\textrm{c}} = $ \qty{10.7(2)}{mJ.cm^{-2}}. Note, that the first skyrmions appear at approximately the same fluence in both samples---similar to the current-driven nucleation---and that the higher threshold fluence in the enhanced stack is the result of an increase in the relative width of the transition region between the labyrinth and skyrmion states~\footnote{To achieve a good signal-to-noise ratio the bias field had to be switched of during the MFM scans, \ie{} after nucleation. This could have affected the relative width of the transition region for the laser-induced nucleation experiment if some of the nucleated skyrmions expand back into worm-like domains at zero field.}. Around the threshold fluence an intermediate region exists for both samples [See \cref{fig:Figure_3}~(h) and (k)]. Here, the state can be characterized by the growth of the skyrmion density in the fluence range from approximately $9$ to $12$ \unit{mJ.cm^{-2}} [see \cref{fig:Figure_3}~(n)]. The appearance of different magnetic textures happens in the intermediate region: bubbles with a light core (parallel to the bias field), worm domains, skyrmionium-like structures, and skyrmions with a black core (antiparallel to the bias field).

After the intermediate region, the skyrmion phase appears, characterized by a saturation of the skyrmion density [\cref{fig:Figure_3}~(n)]. Only one type of magnetic texture is present in the samples in this region: skyrmions with a dark core in a light background, \ie{} the magnetization in the core is antiparallel to the bias field. By comparing the magnetization state in \cref{fig:Figure_3}~(i) and (l) and the two datasets in \cref{fig:Figure_3}~(n), it is clear that the enhancement of the DMI leads to an increase in skyrmion density in the whole fluence range. The skyrmion density in the saturation region in the enhanced sample is $4$ times higher than in the reduced sample, which correlates with the difference in the effective DMI for both samples.

\subsection{Bias field dependence}
\begin{figure}
\includegraphics{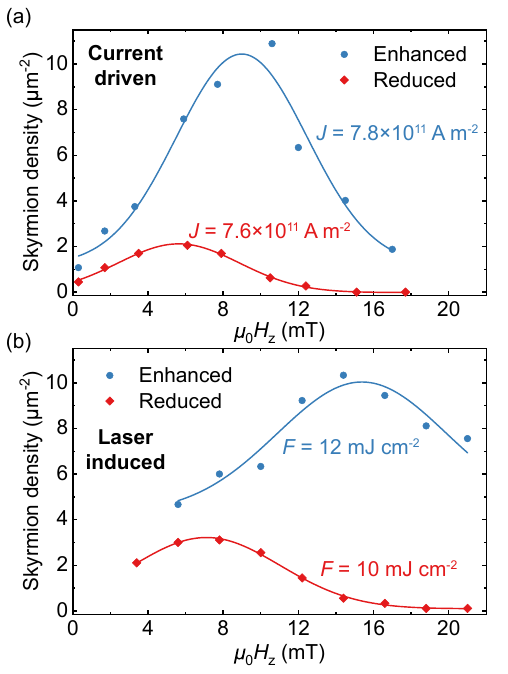}
\caption{
Dependence of the number of skyrmions after nucleation on the strength of the bias field applied during nucleation. 
(a) Current-induced nucleation (b) laser-induced nucleation. The lines are a guide to the eye.
}
\label{fig:Figure_4}
\end{figure}

Finally, we examined the dependence of the number of skyrmions after nucleation on the bias field that is applied during nucleation, for both nucleation methods and stack types. In \cref{fig:Figure_4} we plot this data for the current-driven and laser-induced skyrmion nucleation experiments. For the current density we used the maximum current density that could be reached with our pulse generator, the laser fluence in \cref{fig:Figure_4}~(b) however, is slightly lower than the maximum fluence used in \cref{fig:Figure_3}.

At low bias fields the magnetization in all stacks consists mostly of worm-like domains with an occasional skyrmion. As the field increases, the proportion of skyrmions increases until a maximum is reached. For even greater fields, the device contains only skyrmions after nucleation but the number of skyrmions decreases as the field goes up since the skyrmions become less stable at higher fields~\cite{Buettner2018}. The bias field at which the maximum number of skyrmions is found is greater for the enhanced stack than the reduced stack, for both nucleation methods~\footnote{We note that the optimal bias fields are higher in the case of the laser-driven nucleation, even though the ratio between the optimal bias field for the enhanced and reduced stack is the same for both nucleation methods. We suspect that this is most likely the result of the fluence being closer to the threshold fluence for nucleation, but cannot rule out a small calibration error in one, or both, of the experiments.}. Additionally, we find that the skyrmion density is greater in the enhanced stack than in the reduced stack for all bias fields. Taken together, these observations show that skyrmions are more stable in the enhanced stack compared to the reduced stack. We attribute this increase in stability to the increase in the effective DMI due to the dipolar field.

Finally, in Supplemental Material SIV, we show some preliminary measurements that indicate that the spin-orbit torque (SOT) can still drive domain walls and skyrmions in the enhanced stack, even though the stacking order is inverted in half of the stack. This is expected, since the \ce{Pt} layer---which is likely the primary contributor to the SOT~\cite{Fache2020}---is moved from the bottom to the top of the \ce{Co} layer by this inversion, both the SOT and the chirality of the domain walls change sign. Hence, the direction of motion of any chiral magnetic texture in the \ce{Co} should remain the same. Nevertheless, it would be interesting to study how the SOT-driven motion of domain walls and skyrmions is affected by the inversion of the stacking order in half the stack, but this falls outside the scope of this work.

\section{Discussion and conclusion}
We discuss our results in the context of thermally activated skyrmion nucleation models that have been applied to both current-driven and laser-driven excitation~\cite{Lemesh2018,Buettner2021,Gerlinger2021,Kern2022}. In these approaches, atomistic spin-dynamics simulations~\cite{Eriksson2017} describe a single magnetic layer with short-range exchange and DMI and an \OOP{} anisotropy. Upon heating above $\sim 0.5\,T/T_{\textrm{c}}$ in an \OOP{} field, the system enters a topological fluctuating regime in which topological charges appear and annihilate on pico- to nanosecond timescales~\cite{Buettner2021,Lemesh2018}. During subsequent cooling, a fraction of these excitations becomes trapped, yielding skyrmions whose final density depends on field~\cite{Gerlinger2021}, magnetic parameters (\eg{} anisotropy~\cite{Kern2022} and DMI), and likely the cooling rate.

Two ingredients central to our multilayers are typically absent from such simulations because of computational cost: the multilayer nature and long-range dipolar interactions. Without interlayer dipolar coupling, the dipolar enhancement or reduction of the effective DMI does not occur.

We conjecture that in the high-temperature fluctuating regime, the two stacks should behave similarly because the layer composition is identical and the reduction of $M_{\textrm{s}}$ at elevated temperature~\cite{Buettner2021} is expected to weaken interlayer coupling. During cooling, however, dipolar coupling is reinstated and the effective DMI in the enhanced and reduced stacks diverges, primarily affecting skyrmion stability and survival rather than the onset of nucleation. This is consistent with our observation of comparable nucleation thresholds but different post-excitation skyrmion densities, and it naturally explains why current-driven and laser-driven nucleation show the same qualitative contrast between stacks: although the excitation durations differ strongly (\qty{50}{ns} vs.\ \qty{70}{fs}), the relevant cooling dynamics may be much less different. However, static imaging alone, as employed in this work, cannot resolve such time-dependent dynamics and follow-up work might look into time-resolved methods (\eg\ as in this recent work~\cite{FELIX2025}).

To conclude, we engineered \ce{Ir}/\ce{Co}/\ce{Pt} multilayers with a layer-dependent DMI sign by reversing the trilayer stacking order in one half of the stack. This design aligns (or anti-aligns) the interfacial-DMI effective field with the in-plane component of the dipolar field, producing an effective DMI in the enhanced stack that is $2.5$ times larger than in the reduced stack while keeping $M_{\textrm{s}}$ and uniaxial anisotropy comparable. In skyrmion nucleation devices, both nanosecond current pulses and femtosecond laser pulses yield systematically higher post-excitation skyrmion densities in the enhanced stack, whereas the threshold current density and threshold fluence are found to be similar. 

Taken together, these results indicate that a dipolar-field-modified effective DMI, created by choosing the right stacking order for the heavy metal and magnetic layers, is an effective tool to facilitate both current-driven and laser-induced skyrmion nucleation in magnetic multilayers.

A possible next step is to probe the proposed sequence directly with time-resolved measurements~\cite{FELIX2025}, ideally for laser-driven nucleation to avoid additional spin-orbit-torque contributions from the spin Hall effect in \ce{Pt} and \ce{Ir}~\cite{Buettner2017}. Pump--probe magneto-optical Kerr experiments and reciprocal-space probes such as small-angle X-ray scattering~\cite{Buettner2021,FELIX2025} could potentially establish when the dipolar-field-induced modification of the effective DMI becomes operative. On the theory side, extending atomistic spin-dynamics descriptions to multilayers with long-range dipolar interactions and layer-resolved DMI would provide a quantitative framework for interpreting such data.

\begin{acknowledgements}
This project has received funding from the European Union’s Horizon 2020 research and innovation programme under the Marie Skłodowska-Curie grant agreement No 861300 and the ERC grant agreement No 856638 (3D-MAGiC).. We acknowledge the research program “Materials for the Quantum Age” (QuMat) for financial support under registration number 024.005.006 and the "Research Centre for Integrated Photonics" both part of the Gravitation program financed by the Dutch Ministry of Education, Culture and Science (OCW).n We thank Johan H. Mentink and Rein Liefferink for discussions and input to this work. 
\end{acknowledgements}

\bibliography{References}

\end{document}

\title{Supplemental Material: Facilitating electrical and laser-induced skyrmion nucleation with a dipolar field enhanced effective DMI}

\author{Mark~C.\,H.~de~Jong}
\affiliation{\eindhoven}
\author{Dinar~Khusyainov}
\affiliation{\nijmegen}
\author{Julian~Hintermayr}
\affiliation{\eindhoven}
\author{Bart~Sanders}
\affiliation{\eindhoven}
\author{Dmitry Kozodaev}
\affiliation{\AFM}
\affiliation{\nijmegen}
\author{Aleksei~V.~Kimel}
\affiliation{\nijmegen}
\author{Bert~Koopmans}
\affiliation{\eindhoven}
\author{Theo~H.\,M.~Rasing}
\affiliation{\nijmegen}
\author{Reinoud~Lavrijsen}
\email{r.lavrijsen@tue.nl}
\affiliation{\eindhoven}

\date{\today}

\maketitle

\section{Characterization: SQUID-VSM and MOKE}
\begin{figure}
    \includegraphics[width=\textwidth]{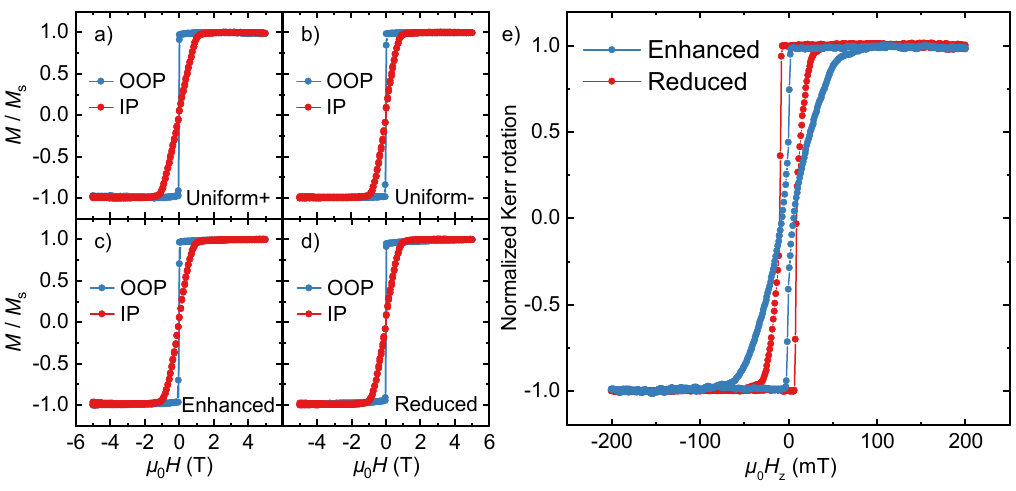}
    \caption{
        a) - d) In-plane and \OOP{} hysteresis loops measured using SQUID-VSM for all four magnetic multilayer studied in the main text.
        e) Response of the sample to low \OOP{} magnetic fields for both the enhanced and the reduced sample, measured using the magneto-optical Kerr effect in polar mode.
        The enhanced sample has a remanence of zero, as domains form before zero applied field is reached due to the low domain-wall energy density, the reduced stack on the other hand has $100 \%$ remanence.
    }
    \label{fig:sup:Alternating_SQUID}
\end{figure}

For each magnetic multilayer studied in the main text we measured \IP{} and \OOP{} hysteresis loops using SQUID-VSM.
The results are plotted in \cref{fig:sup:Alternating_SQUID}~a) through d).
We extract the saturation magnetization $M_{\textrm{s}}$ from the measured moment at large fields, $\lvert \mu_{0} H \rvert > $ \qty{3}{T}, and the effective anisotropy $K_{\textrm{eff}}$ from the area between the loops~\cite{Johnson1996}, the results are summarized in \cref{tab:sup:Alternating_SQUID}.
Additionally, we show the response of the enhanced and reduced sample to an \OOP{} magnetic field sweep measured using polar MOKE in \cref{fig:sup:Alternating_SQUID}.
The enhanced sample is demagnetized due to the formation of a labyrinth domain state at zero field, but the reduced sample has a $100 \%$ remanence at zero field.
This is due to the change in the balance between the dipolar energy, which decreases with domain formation and the domain-wall energy, which increases with domain formation.
Due to the increased effective DMI, the formation of domain walls is less costly in the enhanced stack, resulting in the decreased remanence.

\begin{table}
    \centering
    \caption{
        Magnetic parameters extracted from the hysteresis loops measured using SQUID-VSM shown in \cref{fig:sup:Alternating_SQUID}.
        And the equilibrium domain width extracted from the MFM scans in \cref{fig:sup:Alternating_Demag}, as well as, the corresponding effective DMI strength calculated using the model of Lemesh (assuming that $A = $ \qty{12}{pJ.m^{-1}}).
    }
    \label{tab:sup:Alternating_SQUID}
    \vspace{1mm}
    \begin{tabular}{l c c c c c}
        \hline\hline
        Device  &   $M_{\textrm{s}}$ (\unit{MA.m^{-1}}) &   $K_{\textrm{eff}}$ (\unit{MJ.m^{-3}}) &   $K_{\textrm{u}}$ (\unit{MJ.m^{-3}}) &   $W$ (\unit{nm})    &   $D_{\textrm{Eff}}$ (\unit{mJ.m^{-2}})\\
        \hline
        Uniform$+$  &   $0.92\pm0.03$   &   $0.42\pm0.01$  &   $0.96\pm0.03$    &   $230\pm3$    &   $1.5\pm0.1$\\
        Uniform$-$  &   $1.25\pm0.01$   &   $0.426\pm0.004$  &   $1.41\pm0.02$  &   $400\pm12$   &   $0.89\pm0.04$\\
        Enhanced    &   $1,11\pm0.02$   &   $0.421\pm0.009$  &   $1.21\pm0.03$  &   $177\pm3$    &   $1.54\pm0.06$\\
        Reduced     &   $1.14\pm0.03$   &   $0.42\pm0.01$  &   $1.24\pm0.04$    &   $416\pm8$    &   $0.6\pm0.1$\\
        \hline
    \end{tabular}
\end{table}

\newpage
\section{Characterization: MFM}
\begin{figure}
    \includegraphics{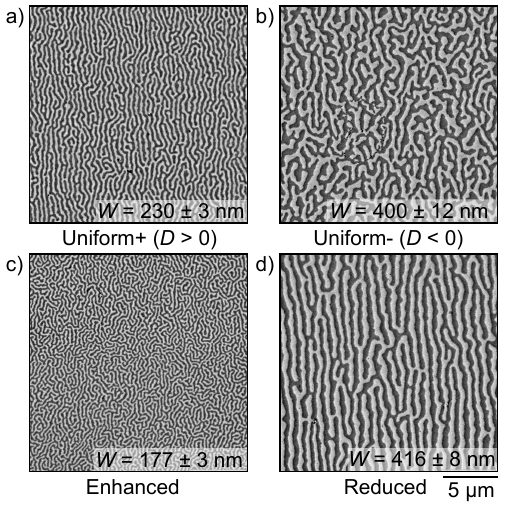}
    \caption{
        The domain state after demagnetization for a) the uniform$+$ stack, b) the uniform$-$ stack, c) the enhanced stack, and d) the reduced stacks.
        Each scan has a size of $20 \times 20$ \unit{\micro\metre\squared}.
        The inset shows the equilibrium domain size $W$ that is extracted using the Fast-Fourier Transform of the MFM data.
    }
    \label{fig:sup:Alternating_Demag}
\end{figure}

In \cref{fig:sup:Alternating_Demag}~a) through d), we plot the domain state that was measured using MFM in each of the four magnetic multilayers studied in the main text.
Each sample was demagnetized using an oscillating magnetic field at an angle of \qty{85}{\degree} away from the sample normal.
The initial field strength was $\mu_{0} H = $ \qty{5}{T} and the next field strength was given by $\mu_{0} H_{\textrm{next}} = -0.995 \times \mu_{0} H_{\textrm{previous}}$ until a field of $\mu_{0} H = $ \qty{5}{mT} was reached.
The sweep rate of the field was set to \qty{50}{mT.s^{-1}}.
We set the scan size for each MFM scan to $20 \times 20$ \unit{\micro\metre\squared}---to ensure that enough domains are present in the scan to accurately determine the domain width from the Fast-Fourier Transform of the scan---and the lift height was set to \qty{30}{nm}.
In \cref{tab:sup:Alternating_SQUID}, we summarize the results and also give the effective DMI strength calculated using the model of Lemesh~\cite{Lemesh2017}.

\newpage
\section{Skyrmion nucleation sites and skyrmion size}
\begin{figure}
    \includegraphics{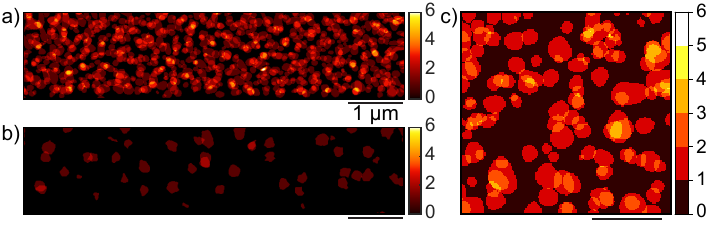}
    \caption{
        Cumulative binarized MFM scans taken above the nucleation threshold, pixels where a skyrmions is present contribute $1$ to the total signal.    
        a) Current-driven nucleation in the enhanced stack, six scans.
        b) Current-driven nucleation in the reduced stack, six scans.
        c) Laser-induced nucleation in the reduced stack, 5 scans.
    }
    \label{fig:sup:Alternating_Locs}
\end{figure}

To determine if the skyrmions are nucleated at specific positions---such as pinning sites---or at random locations in the devices we plot the cumulative binarized MFM scans taken above the nucleation threshold. The cumulative images in \cref{fig:sup:Alternating_Locs}~a) and b), correspond to the current-driven nucleation in the enhanced and reduced stack, respectively, and each contain the six binarized MFM scans with the largest current density. Hence, the color of each pixel corresponds to the number of skyrmions that were detected at that location. In both cases, we observe that the nucleation is spread out over the entire device and does not localize at specific positions, which corresponds well with other reports on temperature driven nucleation in the literature~\cite{Legrand2017,Lemesh2018,deJong2023}. For the laser-induced nucleation the corresponding dataset is limited, as we only performed repeated nucleation for a single fluence on the reduced stack. The cumulative binarized image consisting of five individual nucleation events is shown in \cref{fig:sup:Alternating_Locs}~c). As was the case for the current-driven nucleation on the reduced stack, we again do not observe a preferred nucleation site and the skyrmions appear to be randomly distributed over the scanned area. This is also consistent with the behavior reported in literature~\cite{Buettner2021,Gerlinger2021,Kern2022}.

In addition to the nucleation locations, we examined the average skyrmion size in our enhanced and reduced magnetic multilayers. Because the magnetic field had to be switched off to reduce the noise in the MFM scans in the laser-induced nucleation experiments, we only report the skyrmion size for the current-driven nucleation experiments---since these were measured when the bias field was applied. The lift height for the backward pass was set to \qty{40}{nm} for all scans discussed in this section. To determine the skyrmion size we first identify which domains should be counted as skyrmions. To do this, we use the same skyrmion identification script used in Ref. \cite{deJong2023}---however, the threshold parameters are tweaked slightly to optimize the script on the new dataset. The skyrmion diameter is determined using the \verb|MajorAxisLength| property as determined by the \verb|regionprops()| function in MATLAB. Note that---because the signal measured in MFM is a convolution of the tip magnetization and the stray field of the sample---this overestimates the size of the skyrmions~\cite{Kazakova2019,Barton2023}.

\begin{figure}
    \includegraphics{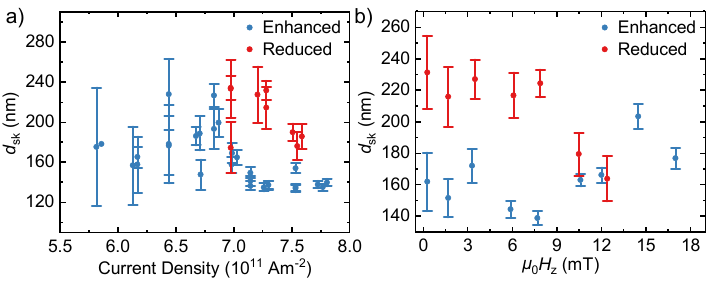}
    \caption{
        Average skyrmion size for all MFM scans as a function of a) current density and b) the applied bias field.
        This figure only contains MFM scans taken after current-induced skyrmion nucleation.
        The error bars represent the standard error of the average.
    }
    \label{fig:sup:Alternating_Skyrm_size}
\end{figure}

In \cref{fig:sup:Alternating_Skyrm_size}, we plot the extracted skyrmion diameter $d_{\textrm{sk}}$ as a function of the applied current density for all MFM scans used to create Fig. 3~m) in the main text.
Below the critical current density ($J_{\textrm{c}} = $ \qty{7.09(1)e11}{A.m^{-2}}) we observe that the skyrmion size has a large variance, due to the low number of skyrmions in the device.
When the critical current density is crossed the skyrmion size converges to the same value, independent of the current density that is applied, which we take as the skyrmion diameter---this behavior is especially clear for the enhanced stack which has a much higher number of skyrmions in the final MFM images.
We find $d_{\textrm{sk}} = $\qty{135(3)}{nm} for the enhanced stack and $d_{\textrm{sk}} = $ \qty{184(7)}{nm} for the reduced stack.

We can also extract the skyrmion diameter for the field dependent measurements shown in Fig. 4~a) in the main text.
However, in the case of the enhanced stack, the AFM tip was damaged for the scans taken for fields $\mu_{0} H > $ \qty{10.6}{mT}, this affected the measured skyrmion size but not the skyrmion density discussed in the main text.
For the reduced stack we observe that for low fields the skyrmion size is approximately constant, then for fields larger than \qty{9}{mT} the skyrmion size starts to decrease until skyrmions are no longer stable.
The constant skyrmion size at lower fields is likely the result of pinning of the domain walls, as the skyrmions are quite isolated in the device and relatively small compared to the width of the strip.
A similar behavior is expected for the enhanced strip, but unfortunately we cannot draw any conclusions about the skyrmions size due to the broken tip.
\newpage
\section{Preliminary experimental results on SOT-driven motion}
\begin{figure}
    \includegraphics{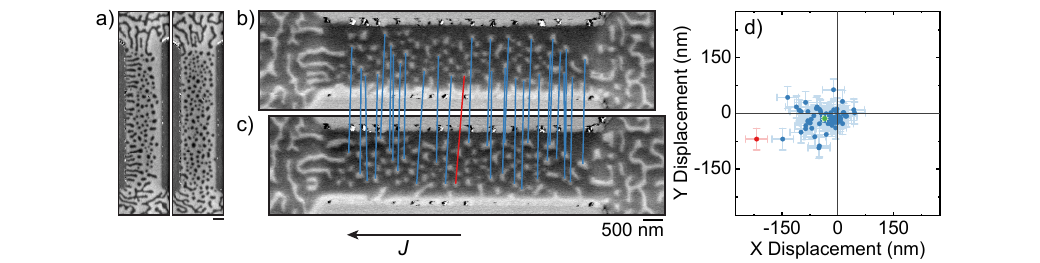}
    \caption{
        a) Domain states in the enhanced stack after the application of a single current pulse, these scans are also shown in Fig. 3~(b, c) of the main text.
        A nucleation device before b) and after c) the application of $11$ current pulses (length: \qty{1}{ns}; magnitude: \qty{1e12}{A.m^{-2}}).
        The defects on the edges of the device were used to align the two scans.
        The blue and red lines indicate the position of several skyrmions before and after the pulses.
        In d), we plot the displacement as measured for all $74$ skyrmions in the MFM scan.
        On average the skyrmions move to the left and down, the average displacement of all skyrmions is denoted by the green dot---the error bars represent the standard error of the average.
        The red line and data point correspond to the skyrmion for which the maximum displacement was observed.
        All scale bars are \qty{500}{nm}.
    }
    \label{fig:sup:Alternating_Sk_Motion}
\end{figure}

Here, we show two preliminary analyses that reveals that skyrmions are driven by SOT in our enhanced magnetic multilayer.
In \cref{fig:sup:Alternating_Sk_Motion}~a), we show the MFM scans of the domain state that is nucleated using current-driven nucleation in Fig. 3~(b, c) of the main text, where an effect of the current can be seen in the funnel shaped regions above and below the strip in.
In the top funnel, all the dark domains are oriented vertically---parallel to the strip---while in the bottom funnel the dark domains are oriented horizontally---perpendicular to the strip.
Although we only show two scans here, this phenomenon was observed for all scan taken for $J \geq $ \qty{5.86e11}{A.m^{-2}} on the enhanced stack.
We hypothesize that this is a consequence of spin-orbit torque (SOT) driven domain-wall motion.
As the current flows into the strip in the top funnel, the current density gradually increases until at some point a critical current is reached and the domain walls are pulled towards the strip, elongating the domains in the direction parallel to the strip.
In the other funnel the opposite effect occurs, here the current density gradually decreases, which causes the moving domains to bunch up.
In this scenario, the domain walls move along the current flow, and against the electron flow, which is consistent with a spin-orbit torque (SOT) due to the spin-Hall effect in \ce{Pt} causing the domain wall motion~\cite{Haazen2013,Emori2013}.

Additionally, we tried to use \qty{1}{ns} current pulses with a high current density, $J = $ \qty{1e12}{A.m^{-2}}, to move the skyrmions in a nominally identical device.
We nucleate skyrmions in a nominally identical device to that used in the main text, using a bipolar pulse train.
The pulse length is \qty{10}{ns} and the magnitude is $J = $ \qty{7.1e11}{A.m^{-2}}, the first pulse flows from right to left and \textit{vice-versa} for the second pulse.
The bias field is the same as in the main text, $\mu_{0} H = $  \qty{10.3(5)}{mT}.
The resulting skyrmion state is shown in \cref{fig:sup:Alternating_Sk_Motion}~a).
We then use a second pulse generator (Picosecond Pulse Labs model A10,070A) to send a train of $11$ current pulses with a length of \qty{1}{ns} and a magnitude of $J = $ \qty{1.0e12}{A.m^{-2}} through the device, all these pulses flow from right to left (electrons move from left to right).

In \cref{fig:sup:Alternating_Sk_Motion}~b), we show the state of the magnetization after the application of these pulses, the image has been shifted to the right to align the defects that are present on the device edges in both scans.
We then compare the position of each skyrmion before and after the pulses, in \cref{fig:sup:Alternating_Sk_Motion} we show the change in position for a subset of the skyrmions using the blue and red lines.
The horizontal displacement is small, but clearly not zero for some skyrmions.
To see this in more detail, we extract the position of each skyrmion manually from both scans, the displacement for each skyrmion is plotted in \cref{fig:sup:Alternating_Sk_Motion}~c).
The uncertainty represents a $100\%$ uncertainty interval for each point corresponding to a region of $6\times6$ pixels in the MFM scan.

The average displacement for all skyrmions is \qty{38(9)}{nm} with a skyrmion Hall angle of $\lvert \theta_{\textrm{sk}} \rvert = $ \qty{22(5)}{\degree}---as shown by the green dot in \cref{fig:sup:Alternating_Sk_Motion}~c).
This corresponds to an average skyrmion velocity of $v_\textrm{sk} = $ \qty{3.4(9)}{m.s^{-2}}.
Furthermore, the horizontal displacement is against the electron flow direction, confirming that the skyrmions are driven by a SOT generated from the \ce{Pt} layers~\cite{Haazen2013,Emori2013}.
The maximum displacement that we observe corresponds to a skyrmion velocity of $v_\textrm{sk} = $ \qty{21(5)}{m.s^{-2}}, and is highlighted in red in \cref{fig:sup:Alternating_Sk_Motion}.
We expect that both pinning and the relatively high density of skyrmions in the narrow strip affect the motion of skyrmions in our stacks~\cite{Legrand2017,Tan2021}, which could explain the relatively large variance in skyrmion displacements observed in \cref{fig:sup:Alternating_Sk_Motion}~c).
Our preliminary results are consistent with the work of Tan \ea{} in Ref. \cite{Tan2021}, who performed a comprehensive statistical analysis of skyrmion motion of dense skyrmion arrays in similar devices fabricated from a [\ce{Pt}$(3)\,|\,$\ce{Co}$(1.2)\,|\,$\ce{MgO}$(1.5)$]$_{\times 15}$ multilayer.
The average skyrmion velocities reported in Ref. \cite{Tan2021} are similar compared to our preliminary findings, although at half the applied current density.
We expect that this is the result of the thicker \ce{Pt}$(3)$ layers used in Ref. \cite{Tan2021}, which increases the spin accumulation generated by the spin Hall effect~\cite{Yin2017,LegrandPhD2019} and by differences in pinning strength caused by the different stack compositions.
The skyrmion Hall angle we observe is also consistent in magnitude and sign with earlier reports~\cite{Litzius2016,Tan2021}, if we assume that \ce{Pt} is the dominant source of the SOT.

Although these results are preliminary---due to the limited number of skyrmions observed and the single device that was tested and the manual estimation of their center pixel in the scan---they are in good agreement with the comprehensive statistical analysis of skyrmion motion in similar devices fabricated from a [\ce{Pt}$(3)\,|\,$\ce{Co}$(1.2)\,|\,$\ce{MgO}$(1.5)$]$_{\times 15}$ multilayer in Ref. \cite{Tan2021}.
Taken together, the observed current-driven motion of domain walls and skyrmions shows that the SOT can still be used to drive chiral magnetic textures in our stacks, even though the stacking order is inverted in half of the stack.
Indeed, since the \ce{Pt} layer---which is likely the primary contributor to the SOT~\cite{Fache2020}---is moved from the bottom to the top of the \ce{Co} layer by this inversion, both the SOT and the chirality of the domain walls change sign.
Hence, the direction of motion of any chiral magnetic texture in the \ce{Co} should remain the same.
\newpage
\section{\mumax{} simulations and modelling}
\noindent\textbf{Micromagnetic simulations}

The micromagnetic \mumax{} simulation~\cite{Vansteenkiste2014} presented in Fig. 2~(d) of the main text is discussed first.
The material system we simulate is $[\textrm{NM}(2)/\textrm{FM}(1)]\times N$, where $N \in [2,4,6,8,10]$ is the total number of repeats of the \qty{2}{nm} thick nonmagnetic (NM) layer and \qty{1}{nm} thick ferromagnetic (FM) layer.
The non-magnetic layers are represented by empty cells, such that the only coupling between the layers is the dipolar coupling.
The magnetic parameters are set equal to the average magnetic parameters of the uniform$+$ and uniform$-$ stacks, as shown in Fig. 2 of the main text: $M_{\textrm{s}} = $ \qty{1.1}{MA.m^{-1}}, $A = $ \qty{12}{pJ.m^{-1}}, and $K_{\textrm{u}} = $ \qty{1.2}{MJ.m^{-3}}.
The DMI strength is varied between $\lvert D_{\textrm{Eff}} \rvert = $ \qty{1}{mJ.m^{-2}} and \qty{3}{mJ.m^{-2}}.
To represent the four magnetic multilayers studied in the main text we control the DMI sign in each half of the stack.

We simulate a region of $512 \times 32$ \unit{nm^{2}} with periodic boundary conditions in the $x$ and $y$-direction of $32$ repeats.
The cell sizes are given by $(x,\,y,\,z) = (0.5,\,8,\,1)$ \unit{nm}.
Two domain walls are then initialized in these systems, with the domain-wall normal along the $x$-direction.
Their positions are $1/4$ and $3/4$ of the simulation region in the $x$-direction, such that the periodic boundary conditions result in a periodic array of equally sized up and down domains.
The width of these initial walls is set to \qty{5}{nm} and the domain-wall moment is set to $45^{\circ}$ from the domain-wall normal.
The systems are subsequently relaxed to find the equilibrium domain-wall moment for each layer.

\begin{figure}
    \includegraphics{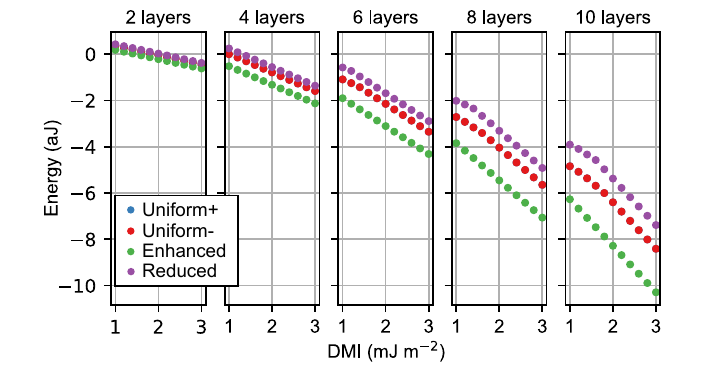}
    \caption{
        Energy difference between a state with two domain walls and no domain walls as a function of the interfacial DMI strength for the four magnetic multilayers studied in the main text---calculated using \mumax{}.
        The data for the two uniform stacks overlap, since the average magnetic parameters from Fig. 2 were used.
        Hence, the only difference between the two uniform stacks is the sign of the interfacial DMI.
    }
    \label{fig:sup:Alternating_MuMax_Sim}
\end{figure}

In \cref{fig:sup:Alternating_MuMax_Sim}, we plot the energy for all four magnetic multilayers as a function of the interfacial DMI strength and the number of layers.
This energy plotted here is the total energy of the system with the two domain walls, minus the energy of the same system in a uniformly magnetized state---which is constant as a function of the DMI and does not affect the results.
To calculate the contribution of the dipolar interaction between the magnetization in the domains and the domain walls to the effective DMI, we interpolate the result and then calculate the difference in interfacial DMI needed to achieve the same energy in the reduced and the enhanced system.
This difference is equal to two times the enhancement / reduction in the effective DMI.

There are two important caveats:
(\textbf{I}) The increase and decrease in the effective DMI is not symmetric with respect to the two uniform stacks.
The reason for this is that both the enhanced and reduced stack have an additional energy gain due to a dipolar coupling between the antiparallel domain-wall moments in the top and bottom half of the stack---compared to the two uniform stacks, where these domain-wall moments are parallel~\cite{Lucassen2020a}.
(\textbf{II}) For the systems with $N \geq 6$, the energy in the reduced and uniform stacks is no longer linear for small interfacial DMI strengths.
This is because the domain-wall moment in one or more of the layers starts to re-orient towards the dipole field of the domains.
In the enhanced stack this never happens, because the dipole field of the domains and the DMI stabilize the same orientation of the domain-wall moment.
To avoid an effect of this on the calculated effective DMI contribution we ensure that this effective contribution is calculated using a set of simulations where the energy is a linear function of the DMI.
Note that, in the four magnetic multilayers that were explored in the main text, the strength of the effective DMI in the reduced stack was only $\lvert D \rvert = $ \qty{0.6(1)}{mJ.m^{-2}}.
According to these simulations, that means that not all the layers are aligned with the orientation favored by the DMI.
In this region of the plot, the slope of the energy vs. DMI curve is lower for the reduced and uniform stacks, which means that the reduction in the effective DMI of the reduced stack compared to the uniform stacks is increased, compared to the increase in the effective DMI of the enhanced stack.
This is consistent with the slightly larger reduction in the effective DMI that we measured for the reduced stack.\\
\\
\noindent\textbf{Analytical calculation of the effective DMI}

To further validate the contribution of the dipole-dipole coupling to the effective DMI that we extract from the \mumax{} simulations we used an analytical model by Lemesh \ea{}~\cite{Lemesh2018a} to estimate this contribution to the effective DMI.
This model is an extension of the model that we use to extract the effective DMI from measurements of the stripe domain width, also by Lemesh \ea{}~\cite{Lemesh2017}.
Below we have reproduced the relevant expressions from Ref. \cite{Lemesh2018a} and mentioned the relevant assumptions that are made---see Refs. \cite{Lemesh2017,Lemesh2018a} for a detailed derivation of the equations.

The system that is considered in the model is a magnetic multilayer consisting of $\mathcal{N}$ magnetic layers of thickness $\mathcal{T}$ and repeat thickness $\mathcal{P}$, extending a distance $L_{x}$ and $L_{y}$ in the $x$- and $y$-direction, respectively.
In each layer the magnetization pattern is identical and consists of stripes oriented along the $y$-direction---\ie{} the magnetization is periodic in the $x$-direction.

Lemesh \ea{} then show that, in this case, the energy density $\sigma_{\textrm{d,sv}}$, corresponding to the interaction of magnetostatic surface charges (generated by the domains) and volume charges (inside the Néel type domain walls) reduces to the following equation,
\begin{equation}
    \sigma_{d,\textrm{sv}} = \mp \frac{\pi \mathcal{T}}{\mathcal{N} \mathcal{P}} \, \sum_{i = 0}^{\mathcal{N} - 1} \, \sin(\psi_{i}) \, D_{\textrm{sv},i}(\Delta_{0}, \dots, \Delta_{\mathcal{N} - 1}),
\end{equation}
\noindent where $\psi_{i}$ and $\Delta_{i}$ are the layer dependent domain-wall angle and domain-wall width, respectively.
The sign of the energy density is determined by the type of domain wall $-$ for a $\downarrow\uparrow$ domain wall and $+$ for a $\uparrow\downarrow$ domain wall\footnote{This is needed because the domain-wall angle $\psi_{i}$ is defined with respect to the Cartesian co-ordinate system and not relative to the domain wall normal---which is opposite for the two domain wall types.}.

In the same notation the energy density corresponding to the interfacial DMI is written as,
\begin{equation}
    \sigma_{\textrm{DMI}}^{1,\mathcal{N}} = \mp \frac{\pi \mathcal{T}}{\mathcal{N} \mathcal{P}} \, \sum_{i = 0}^{\mathcal{N} - 1} \, \sin(\psi_{i}) \, D.
\end{equation}
By inspection, it is clear that the effect of the interaction between the surface and volume magnetic charges is identical to that of the interfacial DMI---albeit with a layer dependent DMI strength $D_{\textrm{sv}, i}$.

$D_{\textrm{sv}, i}$ is layer dependent because its magnitude is determine by the amount of layers above and below the current layer $i$.
It is given by,
\begin{equation}\label{eq:Alternating_ld_DMI}
    D_{\textrm{sv}, i} = -\frac{\mathcal{N}}{\pi f} \, \sum_{j=0}^{\mathcal{N} - 1} \, F_{\textrm{sv}, ij}(\Delta) \, \textrm{sgn}(i - j),
\end{equation}
\noindent where $F_{\textrm{sv}, ij}(\Delta)$ is a matrix whose elements specify the interaction between the surface charge density in layer $j$ acting on the volume charges in layer $i$ and the sgn function keeps track of the position of layer $j$ relative to layer $i$ in the multilayer.
Note, that the layer dependent $\Delta_{i}$ has now been replaced with the layer-averaged domain-wall width $\Delta$, a simplification that is made in Ref. \cite{Lemesh2018a} to reduce the number of free parameters and make the model tractable.
For a system with the magnetic parameters that we measured for our multilayers this assumption should also be valid according to figure 1(d) in Ref. \cite{Lemesh2018a}---which we also verified using the \mumax{} simulations.
The matrix $F_{\textrm{sv}, ij}(\Delta)$ is given by,
\begin{align}\label{eq:Alternating_coupl_mat}
    \nonumber F_{\textrm{sv}, ij}(\Delta) = \frac{\pi \mu_{0} M_{\textrm{s}}^{2} \Delta^{2}}{\mathcal{N} \mathcal{P}} \, \Big[ G(\frac{\lvert (i - j) \mathcal{P} + \mathcal{T} \rvert}{2 \pi \Delta}) + G(\frac{\lvert (i - j) \mathcal{P} - \mathcal{T} \rvert}{2 \pi \Delta})\\ 
    \\
    \nonumber + G(\frac{\lvert (i - j) \mathcal{P} \rvert}{2 \pi \Delta}) \Big],
\end{align}
\noindent where $G(x)$ is,
\begin{equation}\label{eq:Alternating_gamma}
    G(x) = 2 \ln\Big[ \Gamma\Big( x + \frac{1}{2} \Big) \Big],
\end{equation}
\noindent with $\Gamma(x)$ the gamma function.
Using \cref{eq:Alternating_ld_DMI,eq:Alternating_coupl_mat,eq:Alternating_gamma} the layer dependent effective DMI due to the dipolar coupling between domain-walls and domains can be calculated if the domain-wall width is known.

We estimate the average domain wall width using equation 20 from Ref. \cite{Lemesh2017}---using the effective medium approximation to scale the magnetic parameters and while assuming that all the domain walls are fully of the Néel type ($\psi = 90^{o} \vee \psi = -90^{o}$),
\begin{equation}
    \Delta = \Delta_{0} - \frac{1}{\frac{2 \pi (Q - 1)}{\mathcal{NP}} + \frac{1}{\Delta_{0} - \Delta_{\infty}}},
\end{equation}
\noindent with $Q$, $\Delta_{0}$, and $\Delta_{\infty}$ given by,
\begin{align}
    Q = \frac{2 f K_{\textrm{u}}}{\mu_{0} f^{2} M_{\textrm{s}}^{2}},\\
    \nonumber\\
    \Delta_{0} = \sqrt{\frac{f A}{f K_{\textrm{u}}}},\\
    \nonumber\\
    \Delta_{\infty} = \sqrt{\frac{f A}{f K_{\textrm{u}} + 0.5 \mu_{0} f^{2} M_{\textrm{s}}^{2}}}.
\end{align}

\begin{figure}
    \includegraphics{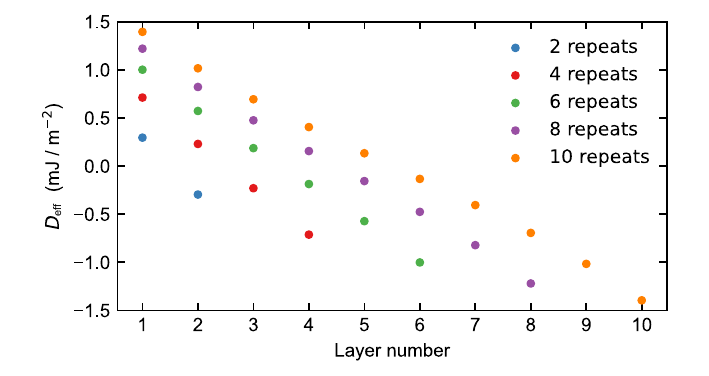}
    \caption{
        Layer dependent contribution of the dipole-dipole coupling between the magnetization in the domains and the domain walls to the effective DMI, plotted for different number of repeats.
        The magnetic parameters correspond to the average magnetic parameters of the uniform$+$ and uniform$-$ stacks reported in Fig. 2 of the main text.
    }
    \label{fig:sup:Alternating_Analytic}
\end{figure}

In \cref{fig:sup:Alternating_Analytic}, we plot the layer dependent contribution to the effective DMI for several magnetic multilayers with different numbers of repeats and the same magnetic parameters as were used in the \mumax{} simulations.
Layers closer to the center of the multilayer gain a smaller contribution to the effective DMI from this effect and the sign of the contribution is opposite for each half of the multilayer---as expected.
In the enhanced stack, this contribution will always add to the effective DMI, and hence the average increase in the effective DMI in the enhanced stack is given by,
\begin{equation}
    \delta D_{\textrm{eff}} = \frac{1}{N} \, \sum_{i=1}^{\mathcal{N}-1} \, \lvert D_{\textrm{sv}, i} \rvert,
\end{equation}
\noindent and in the reduced stack it is $-\delta D_{\textrm{eff}}$.
These are the values plotted in Fig. 2~(d) in the main text, the agreement between the analytical model and the numerical \mumax{} simulations is excellent, which demonstrates that the quantity we extracted from the \mumax{} simulations is indeed the increase or decrease in the effective DMI due to the dipolar field of the domains.

\bibliography{References}